# The Relationship between Centaurs and Jupiter Family Comets with Implications for K-Pg-type Impacts

## K. R. Grazier[1*†], J. Horner[2], J. C. Castillo-Rogez[3]

[1]United States Military Academy, West Point, NY, United States

[2]Centre for Astrophysics, University of Southern Queensland, Toowoomba, Queensland 4350, Australia

[3]Jet Propulsion Laboratory, California Institute of Technology, Pasadena, CA, United States.

[*]Corresponding Author. E-mail: kevin_grazier@yahoo.com

[†]Now at NASA/Marshall Space Flight Center, Huntsville, AL, United States

Centaurs—icy bodies orbiting beyond Jupiter and interior to Neptune—are believed to be dynamically related to Jupiter Family Comets (JFCs), which have aphelia near Jupiter's orbit, and perihelia in the inner Solar System. Previous dynamical simulations have recreated the Centaur/JFC conversion, but the mechanism behind that process remains poorly described. We have performed a numerical simulation of Centaur analogues that recreates this process, generating a dataset detailing over 2.6 million close planet/planetesimal interactions. We explore scenarios stored within that database and, from those, describe the mechanism by which Centaur objects are converted into JFCs. Because many JFCs have perihelia in the terrestrial planet region, and since Centaurs are constantly resupplied from the Scattered Disk and other reservoirs, the JFCs are an ever-present impact threat.

## 1. Introduction

Over the past decade, a number of studies have brought into question the long-held belief that Jupiter acts to shield the Earth from comet impacts. The work of Wetherill (1994, 1995), who studied the influence of the giant planets in clearing debris from the outer Solar System, is often heralded as the source of the "Jupiter: the Shield" paradigm, and was one of the core tenets of the Rare Earth hypothesis of Ward & Brownlee (2000) who popularized the notion.

Grazier (2016), hereafter G16, revisited Wetherill's work with modern, and orders of magnitude more accurate, numerical methods by simulating the trajectories of 10,000 particles initially situated in the Jupiter/Saturn, Saturn/Uranus, and Uranus/Neptune inter-plant gaps. This and other recent studies (e.g., Horner and Jones, 2008, 2009, 2010; and Lewis et al. 2013), revealed the story to be significantly more complicated than previously thought. One key outcome of those studies was the confirmation that, rather than acting as an impenetrable shield, Jupiter acts to increase the flux of Earth-threatening asteroids and short-period comets. This is the result of the dual nature of the planet's influence: in addition to accreting objects, or ejecting them from the Solar System entirely, Jupiter can also hurl them into the inner Solar System.

One of the mechanisms by which Jupiter increases the terrestrial impact flux is by converting Centaurs into Jupiter Family Comets (JFCs). Centaurs—planetesimals with perihelia exterior to the orbit of Jupiter and aphelia interior to the orbit of Neptune—are widely held to be source of the JFCs (e.g. Volk and Malhotra, 2008; Horner, Evans & Bailey, 2004; Levison and *Duncan, 1997)*. JFCs are low-inclination comets with orbital periods under 20 years, many of which have aphelia near Jupiter and perihelia in the terrestrial planet region. The main source region of both JFCs and Centaurs is believed to be the Scattered Disk—a belt of planetesimals with semi-major axes between ~30 AU (some would say 33 AU (Volk and Malhotra, 2008)) and ~50 AU, many of which are Neptune-approaching (e.g., Holman and Wisdom, 1993; Duncan and Levison, 1997; Volk and Malhotra, 2008).

Previous numerical studies have shown that Centaurs and Neptune-approaching trans-Neptunian objects can evolve to encounter Jupiter (e.g. Horner et al., 2004; Horner and Jones, 2009; Grazier, 2016). Once delivered to Jupiter's dynamical control, particles can undergo close approaches with Jupiter that radically alter their orbits, placing them on orbits with one of the apses fixed near the orbit of Jupiter. Of most interest are encounter events—that occur with some frequency—where a particle's aphelion is fixed near Jupiter and its perihelion is placed into the Asteroid Belt or terrestrial planet region. Throughout these various studies, there were numerous instances when these simulations recreated the process by which Centaur objects become JFCs. In fact, many studies explored the inter-relation between Centaurs and JFCs (e.g. Levison and Duncan, 1997; Volk and Malhotra, 2008; Bailey and Malhotra, 2009**),** but the mechanism by which this conversion occurs remains to be fully described.

In this paper, as well as in a companion study (Grazier, Castillo & Horner, 2018; hereafter GCH18), we used techniques inspired by Big Data predictive analytics to mine a dataset output by the G16 simulations—but which, prior to now, has been explored only superficially—that contains information describing the details of 2.61 million planet/planetesimal close approach events. In GCH18, we use this information to detail possible planetesimal evolutionary paths in both the late stages of jovian planet formation and the modern Solar System. In this exploration, we use data mined from that dataset to construct a model by which Centaurs are converted to JFCs. Then we discuss how that process places planetesimals on trajectories that make them potential Cretaceous–Paleogene (K-Pg)-level impactors—a process that is ongoing.

## 2. Methods

G16 recreated much of the simulation work performed by Wetherill (1994, 1995) but with modern, and significantly more accurate, numerical methods (Grazier et al, 2005a,b; Grazier, 2016). One component of G16 was a set of simulations of the orbital evolution of 10,000 particle ensembles originating within the Jupiter/Saturn (JS), Saturn/Uranus (SU), and Uranus/Neptune (UN) inter-planet reservoirs for up to 100 My. The particles studied therein, with a broad range of initial inclinations and eccentricities but with perihelia exterior to the orbit of Jupiter and aphelia interior to Neptune, would be Centaur analogues at the onset of the simulations.

The Sun and planets interacted gravitationally in the G16 simulations, while planetesimals were treated as massless test particles influenced by only the Sun and jovian planets. GM values for the Sun and jovian planets were extracted from JPL Ephemeris DE 245, while the masses of the dynamically insignificant terrestrial planets were added to that of the Sun.

To propagate planet and particle trajectories, G16 employed a modified 13[th] order Störmer multistep integration method (Störmer, 1907) that achieves and maintains the error growth limit known as *Brouwer's Law*. Brouwer's Law (Brouwer, 1937) prescribes that, if the accuracy of the integration is dictated solely by the random error incurred by performing calculations using a finite number of decimal or bit places, and not by any source of systematic error, then the error in energy will grow as $t^{1/2}$, where $t$ is the integration time. Correspondingly, the position error of the planets and planetesimals will grow as $t^{3/2}$. For all simulations in G16, the final system energy error after 100 My is $\mathcal{O}$ ($10^{-10}$) or less while the position errors of all jovian planets is not more than $\mathcal{O}$ ($10^{-4}$) (Neptune) and $\mathcal{O}$ ($10^{-3}$) (Jupiter) radians (Grazier et al., 2005a,b; Grazier et al., 1999).

The simulation code employs an adaptation of the modified Stormer integrator that allows the integration time-step to vary to follow the dynamical timescales—like those associated with events where planetesimals pass close enough to a planet that the planet, not the Sun, is the primary influence on the planetesimal's trajectory. That method, and its error growth properties, is detailed in Grazier, Newman & Sharp (2013).

When the close approach code detects that a planetesimal has entered a planet's gravitational sphere of influence (Danby, 1988), it stores heliocentric state vectors for the planetesimal as well as the Sun and planets. When that planetesimal exits the sphere of influence, again, the code stores particle heliocentric state vectors. From the state vectors stored at close approach ingress/egress, particle initial and final orbital elements can be calculated, as can their changes resulting from the encounter. When particles collide with the Sun, a planet, or when they have been ejected from the Solar System, they are removed from the simulation.

G16 reported previously that planet/planetesimal close approaches within the simulations encompassed the same rich variety of complexity as those that have been documented observationally. While some particles were simply accreted by the planets, or ejected from the Solar System entirely, some became temporarily gravitationally bound to the encountered planet. Some of these captures—known as temporary satellite captures, or TSC orbits—lasted decades. Particles were even temporarily captured into orbits around a planet before impacting it in the manner of comet Shoemaker-Levy 9 (e.g. Hammel et al., 1995). These TSCs became the primary focus of the current study.

We employed a novel data analysis approach that was reflective of the predictive analytics process that commercial retailers employ in suggestive marketing or that Hollywood studios use to assess moviegoer demographics beyond the traditional "four quadrant" model (e.g. Labrinidis, A., & Jagadish, 2012; Gandomi and Haider, 2015). A typical process for a dynamical simulation is usually driven by a testable hypothesis—like the presence, absence or importance of a phenomenon. Flags or triggered output may be incorporated into the simulation code—or its output analyzed using techniques like statistical or Fourier analyses—to yield insight into the existence, relevance, or impact of that phenomenon. On the other hand, the predictive analytics process that we employed begins with two things: a large dataset and the assumption that the data contain answers or insights to questions heretofore unasked. This more exploratory approach has proven to be very powerful, revealing new evolutionary pathways for planetesimals, as demonstrated in GCH18 and this study.

We used a series of micro-applications of the scientific method, proceeding much like a forensic investigation: combing through the G16 close approach dataset to reveal correlations and phenomena already extant in the output, then seeking to uncover the meaning—often by way of follow-up data mining passes. For example, one might initiate such a study by extracting variables

*X* and *Y* from a dataset to determine if there is a correlation. If there is no correlation, *X* and *Y* might then be compared with *Z* to determine a correlation. If *X* and *Z* are correlated, then are both correlated to variable *W*? In the case of our close encounter database, our starting point equivalent of *X*, *Y*, and *Z* are changes to particle orbital elements as a result of the encounter—changes to semi-major axis ($\Delta a$), eccentricity ($\Delta e$), and inclination ($\Delta I$)—where *W* is encounter duration.

This method of data analysis does tend to blur the traditional lines between "method" and "results" reporting for the study—with each dive into the database inspired by the previous. What we lead off with in our results, then, is the trail of bread crumbs that starts with the dive into our close encounter database, and results in a model that describes how, through close approaches to Jupiter, Centaur objects become JFCs.

## 3. Close Approach Statistics and Correlations

We first mined our database of close encounters for changes in orbital elements resulting from all encounters, partitioned by planet and zone of origin. We presented a detailed analysis of many of those results in GCH18, and in this study we were initially more interested in correlations between close-approach-induced-changes in orbital elements,

We found no interesting correlations between $\Delta I$ and $\Delta a$ or $\Delta I$ and $\Delta e$. Table 1, however, presents an examination of the correlations between changes in semi-major axes and eccentricities across close approaches. For every planet in every zone, an increase/decrease of semi-major axis tends to be associated with a corresponding increase/decrease in eccentricity. We discuss this geometry more in sections 5 and 6. The difference between the percentages of the encounters where these values are correlated and those where they are anti-correlated is lower for Jupiter than for the other jovian planets.

A related correlation is on display in Table 2, where each entry represents the average duration (in days) for the class of encounter in the corresponding cell in Table 1. The durations for encounters where the resulting $\Delta a$ and $\Delta e$ values are anti-correlated tend to be dramatically longer, on average, than those for which they are correlated.

## 4. Evolutionary Pathways from the Centaurs and SDO to Jupiter and the inner Solar System

We chose to explore the (typically) long-duration encounters where $\Delta a$ and $\Delta e$ values are anti-correlated. One scenario that could be playing out in these instances is when a planetesimal with aphelion at Saturn and perihelion at Jupiter has a close approach to Jupiter and is redirected. That body would initially have a semi-major axis of 7.39 AU, an eccentricity of 0.30, and a period of just over 20 years. If that planetesimal had an encounter with Jupiter that modified its orbit in such a way that the post-encounter aphelion was fixed near Jupiter and its perihelion in the vicinity of Earth's orbit—if it was converted from a Centaur to an Earth-threatening JFC due to that Jupiter close approach—this would be an anti-correlated encounter. The planetesimal semi-major axis would decrease to 3.4 AU, while its eccentricity would increase to 0.68. In the instance where a Centaur's aphelion was near Uranus, and the perihelion at Jupiter (a = 12.2 AU; $e$ = 0.57), and that body became a JFC due to a Jupiter close approach, it would still see a decrease in semi-major axis, and increase in eccentricity—and the changes in $\Delta a$ and $\Delta e$ anti-correlated. This situation would, certainly, not hold for all Centaurs converted to JFCs, but it does provide a good starting point for inquiry.

Averaged over all encounters, the net particle migration in the G16 simulations was outward, towards the outer Solar system. However, that result is not unexpected, since the eventual fate for most Centaur and cometary objects is ejection from the Solar system (e.g. G16). Before evolving to that end state, however, planetesimals can repeatedly migrate inwards and outwards, and GCH18 followed this behavior to detail those evolutionary pathways that permit planetesimals to be handed down to Jupiter from the more distant jovian planets, even from the Scattered Disk. GCH18 also revealed that this process typically requires many close planetary approaches, oftentimes to the same planet, in order for a planetesimal to migrate inwards to Jupiter.

Once particles encounter Jupiter, the simulations reveal that they are often redirected to the inner Solar System. Fig. 1, a subset of Fig. 5b from G16, displays the perihelion versus aphelion for every particle that passed within 1.5 AU over the entire suite of the 100 My full mass simulations. The majority of particles that passed through the inner Solar System in the simulations had orbital periods less than 20 years, and if we apply the traditional definition, these objects would reside on Jupiter Family Comet orbits. The marked similarities between all three panels of Fig. 1 suggest that most of the particles that passed interior to 1.5 AU did so due to a common mechanism, irrespective of their zone of origin,

The "V" shape structure of each panel of Fig. 1, with the apex of each "V" falling in the vicinity of Q = 5 AU immediately suggests that the mechanism that creates JFCs requires a closer approach to Jupiter. The plots, particularly the SU and UN zone plots, shared hints of a second superposed "V" whose apex fell in the vicinity of Q = 10 AU, suggesting that interactions with Saturn likely contribute to the flux of particles through the terrestrial planet region as well. Grazier et al. found similar for bodies from the inter-jovian reservoirs delivered to the outer Asteroid Belt (Grazier, Castillo & Sharp, 2014). This offers a hint that the mechanism that creates Jupiter Family Comets also occurs at Saturn.

## 5. Planetesimal Orbit Modifications Due to Close Approaches

Table 1 shows that, in our simulations, most planet/planetesimal encounters cause $\Delta a$ and $\Delta e$ to change in a correlated way—either with both increasing or both decreasing—and Table 2 reveals that encounters of this nature are typically brief. In essence, for most particles in these simulations, close approaches to Jupiter are of the hyperbolic single-pass variety.

While some of these encounters certainly owed their brevity to a trajectory skirting the periphery of Jupiter's sphere of influence, not surprisingly, Fig. 2 suggests that most of them owe their short durations to a high initial relative encounter velocity. Displayed in Fig. 2, for all jovian planet close approaches, across all simulations, is the total encounter duration versus the relative planet/planetesimal velocity at the beginning of encounter—when the planetesimal initially enters the gravitational sphere of influence (Danby, 1988). There was a general inverse relationship where higher initial relative velocities typically resulted in shorter encounters, which is intuitive. It was also unsurprising that Jupiter encounters spanned a wider range of relative velocities than those for the other planets—this is simply the result of Jupiter being the innermost jovian: the closer an object to the Sun, the faster it moves, as do planetesimals passing nearby.

An example of a hyperbolic single-pass encounter, as well as its influence on the planetesimal's trajectory, is depicted in Fig. 3. This encounter geometry is much like those used by spacecraft navigators for gravity assists. The vector diagram beneath reveals how the inbound ($v_{in}$) and outbound ($v_{out}$) velocity vectors are equal in magnitude in the planetocentric frame due to conservation of energy, but when translated into the heliocentric frame by adding $v_p$ (yielding $v_{HI}$ and $v_{HO}$), the heliocentric velocity vector changes direction, and increases in magnitude—as does the planetesimal's kinetic energy. As a result, the planetesimal's orbit experiences an increase in semi-major axis and eccentricity—and, given the proper geometry, may be boosted into a Solar System escape trajectory. This geometry is likely the "inverse" of the model presented in this paper that causes Centaurs to become JFCs, and is the process that converts a JFC to a Centaur—or can send either into the Scattered Disk—and is discussed in greater detail in GCH18.

Together, Tables 1, 2, and Fig. 3 re-establish the old spacecraft navigator's rule of thumb for gravity assists: "Pass behind to gain; pass ahead to lose." A close hyperbolic flyby to a planet on the side opposite its velocity vector will produce an increase in heliocentric energy, semi-major axis, and eccentricity. A passage ahead of the planet in its orbit produces a decrease in heliocentric energy and semi-major axis.

Figure 2 shows the degree to which Jupiter can capture particles into lengthy encounters— often spanning decades—irrespective of their reservoir or origin in the simulations. Jupiter also pulls particles into long-term encounters over a much wider range of ingress velocities than the other jovian planets as well, and this is another topic discussed in greater detail in GCH18.

We mined the G16 database for encounters where the ingress particle trajectory had a perihelion greater than 5.2 au (Jupiter's semi-major axis distance). Within the selected collection of encounters, we then searched for those encounter where the particle's post encounter aphelion was less than Jupiter's semi-major axis, and its perihelion was less than 3.3 au (the outer boundary of the Asteroid Belt). Although this represents a moderately constrained set of parameters defining a Centaur to JFC conversion, plotted in cyan in panel A of Fig. 2 are 1994 instances that meet these criteria. Included in this number were particles that began the simulations in the JS, SU, and UN reservoirs. This result confirms that Jupiter encounters can create JFCs from Centaurs. Jupiter's ability to capture bodies into long-term captures over a wide range of approach velocities implies that converstions can occur largely independent of initial starting zone and largely decoupled from evolutionary history, and the similarities in all three panels of Fig. 1 point to a common mechanism.

## 6. A Model for Centaur/JFC Conversion

Given the results presented in Table 2 and Fig. 3, it is a reasonable expectation that the encounters that place Centaurs on JFC orbits should have lengthy durations. A lower bound estimate for the duration of a simple planetesimal encounter with Jupiter starts by assuming an initial planet/planetesimal encounter velocity of $3.31 \times 10^{-3}$ AU/Day, which is the average initial velocity calculated across nearly half a million Jupiter encounters. If we assume a straight path through Jupiter's sphere of influence—where $r_{SOI}$ is approximately 0.322 AU, and assuming no acceleration or curvature—the duration of that passage would be just under 195 days. For an upper bound, we take the duration of one complete planetesimal orbit around Jupiter at the periphery of its sphere of influence: just under 2164 days (approximately six years). Fig. 2 reveals that the simulations replicated hundreds of encounters greater than our upper bound, and tens of thousands significantly greater than the lower.

Table 2 shows that the encounter durations when $\Delta a$ and $\Delta e$ are anti-correlated typically span much longer durations than our 195 day lower bound. Often, especially for particles originating in the Saturn/Uranus and Uranus/Neptune reservoirs, the average encounter duration spans years, and is substantially longer than the 2164 day period of a particle orbiting at the periphery of Jupiter's sphere of influence.

Such results reproduce a situation observed for objects encountering Jupiter on several occasions over the last century. Rickman & Halmort (1981) studied the temporary capture of comet 82P/Gehrels 3 by Jupiter through the middle of the 20[th] Century, and Tancredi et al. (1990) studied similar behavior for comet 111P/Helin-Roman-Crockett which was captured by orbited Jupiter in December 1973, spent the next 11 ½ years in a TSC, and will be recaptured in the year 2075. The most dramatic illustration of such a temporary capture event came during the early 1990s, with the disruption of comet Shoemaker-Levy 9 by Jupiter, and its subsequent collision with that planet. In investigating the evolution of the comet prior to its discovery, Chodas and Yeomans (1996) reported that comet Shoemaker-Levy 9, which was tidally disrupted by Jupiter in 1992 (with the fragments impacting that planet in 1994), was likely captured into orbit around that planet in the year 1929 with an uncertainty of ± 9 years.

Although our code has reproduced Shoemaker-Levy 9-like captures with subsequent impacts, the more common scenario is that the captured particle eventually exits Jupiter's sphere of influence after its tenure as a temporary satellite of the giant planet. The durations displayed in Table 1 and Fig. 2, in comparison to our estimated lower and upper bounds, are diagnostic of trajectories with a high degree of deflection—or even long-term TSC orbits—and these longer encounters often produce anti-correlated $\Delta a$ and $\Delta e$ post-encounter.

An encounter producing a decrease in semi-major axis with a corresponding increase in eccentricity becomes more likely when all close approach egress vectors are possible. In the instances of long-duration encounters—with correspondingly large deflections in trajectory—Fig. 4 indicates that what is of prime significance in converting a Centaur into a JFC is the jovocentric orientation of the planetesimal's velocity vector as it exits the planet's gravitational sphere of influence.

Figure 4 shows Jupiter's sphere of influence ingress and egress information for all of the cyan points in Fig. 2, with the three panels representing particles originating in the JS, SU, and UN zones. The reference frame is Sun-centered ecliptic: a rotating reference frame with the Sun always positioned at 0°. Black points represent ingress points for Centaurs that left the encounter as JFC's, and the radial units are in multiples of $r_{SOI}$, the radius of Jupiter's sphere of influence. Using the same criteria as for the Centaur-to-JFC transitions plotted in cyan in Fig. 2, ingress trajectories were constrained to having perihelia exterior to 5.2 AU. The small number of points sunward of the 90° – 270° line represent events where the particle began the encounter near Jupiter's aphelion but, given the constraints on what defines a Centaur/JFC conversion in this instance, it is expected that most egress points are between 90° and 270°. Further, it is expected that most of these points would lie between 90° and 180° as particles, handed down from more distant jovian planets overtake Jupiter near their perihelion—explaining the qualitative similarities in all three panels.

The red vectors represent egress geometry information for encounters that converted Centaurs to JFCs. The length of each red vector represents the number of particles exiting Jupiter's sphere of influence—as the centerline of 10° bins, with the lengths corresponding to the number of particles that left Jupiter's sphere of influence with aphelia Q <- 5.2 AU, and q <= 3.3 AU (normalized to 1.0). The egress plots in all three panels of Fig. 4 are qualitatively similar, which not only implies that egress geometry dictates what high-deflection or temporary capture encounter result in newly-formed JFCs, this similarity in the three plots also explains the similarity of all three panels of Fig. 1.

Based upon the data presented in Fig. 4, Figure 5 depicts a typical Centaur/JFC conversation encounter. The planetesimal overtakes and encounters Jupiter at, or near, perihelion. After a high-deflection encounter, or even several temporary capture orbits, the planetesimal exits Jupiter's sphere of influence with a jovocentric velocity vector anti-parallel to Jupiter's, but with a heliocentric velocity vector parallel to Jupiter's, having $q$ significantly smaller in magnitude than $v_{planet}$.

Similar to Fig. 2, the vector diagram in Fig. 5 reveals how the inbound ($v_{in}$) and outbound ($v_{out}$) velocity vectors are equal in magnitude in the planetocentric frame, but when translated into the heliocentric frame by adding $v_p$ (yielding $v_{HI}$ and $v_{HO}$), the heliocentric velocity vector, orbital kinetic energy and, hence, semi-major axis all decrease dramatically. Depending upon the magnitude and direction of $v_{HO}$, the eccentricity is likely to increase. Irrespective of where the particle entered Jupiter's sphere of influence, if the particle leaves the encounter with $v_{out}$ anti-parallel to $v_{planet}$, the particle would have a heliocentric velocity significantly less than Jupiter's, and would subsequently fall sunwards. The egress point would be the new aphelion—and the particle left, consequently, in a JFC orbit. Clearly, planetesimals would be on JFC orbits for a range of geometries (roughly) centered on the orientation of $v_{out}$, meaning the right hand panel of Fig. 4 depicts an idealized instance of a Centaur/JFC conversion.

The model predicts that conversions will tend to occur after high deflection or TSC encounters, and would have lengthy encounter durations. The average duration for all 677,000 encounters plotted in Panel A of Fig. 2 is 253 days. The average duration for cyan points—where the encounter created a JFC—is 444 days (445 for particles from the JS zone, 454 for SU, and 421 for UN); There were also very lengthy TSCs that became JFCs: for the longest JS particle encounters, the particle was within Jupiter's sphere of influence for 8505 days, or just over 23 years. For SU and UN zone particles, those values are 11950 (32.7 years) and 15431 (42.2 years) respectively.

The conversion geometries depicted in Figs. 4 and 5 suggest that the requisite egress geometry would be more difficult to achieve for particles on retrograde orbits relative to Jupiter. For all events meeting our conversion criteria, 385, or 19.3 percent, were for retrograde orbits (210 of 1047 for JS zone particles, 112 of 627 for SU, 63 out of 320 for UN).

One can envision that if a retrograde Centaur entered the jovian sphere of influence near the egress vector's antipode, it could exit at the proper location and with the proper velocity vector orientation to become a JFC after performing a hyperbolic single-pass encounter. In this scenario, it is a reasonable expectation that retrograde conversions would occur following significantly shorter encounter durations than for prograde conversions. The average duration of all retrograde conversions was 183 days—with the longest duration encounter that created a JFC being 804 days less than $1/10^{th}$ of the longer prograde conversion—while the average prograde conversion was 474 days.

## 7. Converting Centaurs to Jovian Family Comets

The simulations described in G16 suggest that Saturn, like Jupiter, is capable of "grabbing" the aphelion of a particle, and then placing it into an orbit with its aphelion near Saturn and a perihelion interior to Jupiter, perhaps even in the Asteroid Belt or terrestrial planet region. Similar to Fig. 3, Figure 5 shows all instances where particles (black dots) entered Saturn's sphere of influence with q >= 5.2 AU, and Q <= 30 AU, and egress geometry information in 10° bins for all particles that left the encounter with an aphelion Q <= 9.56 AU (Saturn's semi-major axis), and q <= 5.2 AU.

Since the scenarios explored in Fig. 5 are slightly different from those for Fig. 3, the plots appear qualitatively different. In Fig. 3, the definition of what constitutes a Centaur implies that the vast majority of objects approach Jupiter in the anti-sunward direction. In the case of Fig. 5, Centaur objects can approach Saturn's sphere of influence isotropically. The ingress points in the simulations are not isotropic, however, because most points that encountered Saturn in this scenario were boosted out to the vicinity of Saturn by an encounter with Jupiter, were near their aphelia, and were overtaken by Saturn as a result of differential Keplerian motion. This explains why the majority of encounters began in the 270° to 0° quadrant.

Statistics mined in GCH18 revealed that although particles were moving prograde relative to the Sun, roughly half were retrograde in the planetocentric frame. Due to the different range of ingress geometries for Saturn encounters in Fig. 5 compared to those in Fig. 3, more of the encounters in Fig. 5 were retrograde relative to Saturn. This pulled the most common egress geometries for particles in this scenario to higher angles relative to the Saturn-Sun line.

Because Saturn has less than 1/3 the mass of Jupiter it does not attract planetesimals into long-term TSC orbits as readily. Consequently, the encounter durations were correspondingly shorter. The longest encounter that resulted in a particle having its aphelion fixed at Saturn, with a perihelion interior to Jupiter, was approximately 2 years for particles originating in all zones (1.96 years for JS, 2.15 for SU, 2.10 for UN).

If life imitates simulation, this predicts the existence of a collection of "Saturn Family Comets". Table 3 lists 19 such objects from the JPL *Horizons* Database. While two of the objects, with perihelia exterior to Jupiter, would be properly classified as Centaurs, the remainder are the predicted Saturn Family Comets (hereafter SFCs).

The lengthy encounter durations evident in Fig. 2, panels C and D, also suggest that the ice giants are capable of injecting Centaurs into Uranus Family Comet (UFC) and Neptune Family Comet (NFC) orbits. Given the lower masses of these planets, and their greater distance from the inner Solar System, such occurrences are expected to be markedly less frequent than the creation of JFC and SFC objects.

Table 3 displays the orbital properties of outer Solar System objects from the JPL *Horizons* database that have aphelia within one sphere of influence's distance from the orbits of each of the outermost planets, and perihelia interior to the next innermost planet. Very few of these objects have perihelia interior to the Asteroid Belt. In fact, most have perihelia exterior to Jupiter, and would still be classified as Centaurs, and low-inclination bodies that plunge into the inner Solar System, crossing the realm of Jupiter and Saturn, are certain to be perturbed out of these orbits on short timescales. Nevertheless, these bodies may have evolved into their present orbits through the close approach geometry we describe above.

Although Saturn, Uranus, and Neptune Family Comets are not terms in the modern comet classification nomenclature, Wilson (1909) used these teams at the dawn of the 20[th] Century. The usage in that instance referred more to the jovian planet to which a cometary object passed closest, and not to the body that, through dynamical interaction, placed the comet on its current trajectory—typically with one of the apses fixed at that planet's orbit.

The points forming the x-axis tails of increasingly long encounter durations in Fig. 2 span a significantly greater range of initial relative velocities for Jupiter than for the other jovian planets—discussed in greater detail in GCH18. This means that for encounters over a range of initial relative velocities, Jupiter can still capture particles into long-duration orbits: large deflections or temporary captures. This reveals the strength of Jupiter's gravitational influence in comparison to the other jovian planets, and helps to demonstrate the dominant role the giant planet plays in directing cometary bodies to the inner Solar system—as evidenced by the large JFC population.

## 8. The Road to Becoming Potential K-Pg-like Impactors

Duncan and Levison (1997) found that Edgeworth-Kuiper Belt objects initially on Neptune-approaching orbits can evolve into JFCs, and estimated that one of these becomes Earth-threatening roughly every 13 million years. Our result is more general and not confined to objects that are initially Neptune-approaching. Horner et al. (2004) similarly found that some Centaurs can become short-period comets and potential terrestrial planet impactors.

GCH18 concludes that Centaurs and SDOs can interchange dynamical families many times over the age of the Solar System and do not appear to be dynamically distinct populations. GCH18 also details dynamical pathways that allow distant Centaurs and SDOs to migrate into orbits that approach Jupiter and Saturn. Although Alvarez et al. (*1980*) hypothesized that the impact at the K-Pg (then K-T) boundary that led to the extinction of 75% of life on Earth (Jablonski and Chaloner, 1994) was the result of an asteroid impact, Moore and Sharma (2013) instead make the case that a cometary impact triggered Earth's most recent mass extinction. Both Pope et al (1997) and Vickery and Melosh (1990) have argued against the impactor being a long-period comet from the Oort Cloud. Pope et al. (1997) suggested that the K-Pg impactor could have been a carbon- and water-rich short-period comet, and our results reveal several evolutionary paths that suggest the impactor very plausibly could very plausibly have been a Centaur, a Scattered Disk object, a Plutino, a jovian Trojan, even a Classical Disk member of the Edgeworth-Kuiper Belt (GCH18) turned JFC.

If the energy release from the K-Pg impact was $3 \times 10^{23}$ joules, and assuming the impactor was an icy Centaur 13 km in diameter (Collins et al., 2008; Artemieva, and Morgan, 2009) with an assumed density of 1.0 g/cm$^3$, then the impact velocity would have been on the order of 22.9 km/s. Although the impact velocity would, in part, be dependent upon the approach geometry, this velocity is significantly less than the $v_\infty$ of several Earth-approaching meteoroid streams with asteroidal or short-period cometary progenitors. If the impactor was a 10 km icy body, the impact velocity increases to 33.8 km/s, still similar to or less than the $v_\infty$ for several Earth-approaching meteoroid streams. An example is the Geminid shower. Meteoroids from the parent asteroid 3200 Phaeton approach Earth at 33.7 km/s. Short-period comet 8P/Tuttle is the progenitor of the Ursid shower, whose meteoroids approach Earth at 32.9 km/s.

In Fig. 4, including geometric information for those Jupiter encounters that convert Centaurs to JFCs, only 11 objects wound up on Mars-crossing orbits, and, of those, only 6 were Earth-crossers. These are the results of single encounters, however. Apart from converting non-Earth-threatening Centaurs into terrestrial-planet-crossing JFCs, the G16 simulations have revealed various methods by which Jupiter can drive sunwards the perihelion of a planetesimal with a perihelion already in the terrestrial planet region.

The Horner et al. (2004), G16, and GCH18 studies observed that although Jupiter and Saturn with their present masses can fix planetesimal aphelia at their orbital distances, in simulations with jovian cores—again, recreating the work of Wetherill (1994, 1995) who called these "failed Jupiters"—the cores did not create JFCs. What these studies did note was that multiple encounters, even with jovian cores, can deliver planetesimals to the inner solar system through a series of successive hyperbolic gravity-assist-style passes. As with Fig. 1, Fig. 7 appeared originally in G16, and shows simulation results for a series of 100 My simulations where the planetary masses where the mass of the jovian embryonic cores (15 Earth masses for Jupiter and Saturn, 1 Earth mass for Uranus and Neptune (Wetherill, 1994). The plots show the aphelion and perihelion distances for every particle that passed interior to 1.5 AU in the simulations. The tendril-like structures represent successive Jupiter passes, with each encounter driving the perihelion further sunwards. We also explored the evolution of particles undergoing such a rapid series of encounters in greater detail for the full-mass simulations in GCH18.

If Saturn can create SFCs, then it is clear that a Jupiter with 30% of its present mass could create JFCs as well, even given the higher encounter velocities at 5.2 AU. Indeed, hints of the importance of mass are to be found in Horner & Jones (2009), where the flux of material routed into Earth-crossing orbits is strongly influenced by Jupiter's mass. As Jupiter becomes more massive, it eventually becomes capable of injecting objects to JFC orbits. The point at which Jupiter is able to create JFCs might influence the amount and nature of the planetesimals delivered to the Asteroid Belt and terrestrial planets during the late stages of planetary formation.

The G16 dataset revealed another scenario by which Jupiter places Earth in harm's way. Figure 1 displays numerous instances where particles on JFC orbits crossed interior to Earth's orbit—although only 6 Centaurs became Earth-crossing JFCs through single encounters with Jupiter, and none with Saturn. The data mining that produced Figures 4 and 6 were constrained to Centaur-to-JFC conversions, but if we relax those constraints, what we see in the simulations, then, is that Jupiter can literally intercept a body on an outbound trajectory, then redirect it back towards the Sun. In GCH18, we reported 4665 instances of encounters where particles with perihelia exterior to the asteroid belt passed into Jupiter's realm, and their orbits were modified such that their egress perihelia were interior to the outer edge of the Asteroid Belt. In that work, we also identified 492 such encounters involving the planet Saturn

That geometry is depicted in Fig. 8. The particle enters Jupiter's sphere of influence on the sunward side, it is captured into a high-deflection encounter, and exits in a tighter orbit with its aphelion fixed at Jupiter. In encounters such as this, Jupiter drives the perihelia of particles, many already JFCs, sunwards. Although this geometry is a subset of those discussed previously, it occurs with enough frequency in the simulations to warrant special examination.

Encounters of this nature at Saturn are implied by Fig. 6, with the most likely ingress points being between 270° and 0°, and the most likely egress points between 60° and 100° in the sun-centered ecliptic frame. This geometry allows bodies to be handed down for low-velocity encounters with planets closer to the Sun. For particles that have Saturn encounters leaving them with perihelia near in the vicinity of Jupiter, this results in the particles approaching Jupiter in a "tail chase" geometry. Those that encounter Jupiter would do so in the direction anti-parallel to Jupiter's velocity—i.e. at a low relative velocity—and would be easy to capture and redirect.

These results show that, in the conversation regarding what type of object slammed into Earth 66 million years ago, inciting the K-Pg extinction, there is another class of object worth greater consideration. In GCH18, we made the case that Neptune-approaching scattered disk objects, Centaurs, and Jupiter Family comets were dynamically indistinct populations, with planetesimals switching categories many times over 100 My simulations. Jupiter plays a major role in many of these transitions, and the results presented here and in GCH18 argue that a body in a JFC orbit is at least as likely a candidate as an Oort cloud comet.

It is a cosmic irony that the G16 study that generated the dataset analyzed in this work set out to recreate the 1994 work of Wetherill, often trumpeted as the foundation of the "Jupiter the Shield" myth. Instead, our study shows that Jupiter and Saturn are reasonably efficient at turning Centaurs into JFCs and SFCs with perihelia in the Asteroid Belt or terrestrial planet region. Given that these processes are ongoing, we conclude that, far from being a shield, Jupiter "targets" Earth and the terrestrial planets by placing non-Earth-threatening Centaurs into short-period orbits where they have frequent opportunities to impact terrestrial planets. In short, Centaurs and Neptune-approaching Scattered Disk objects—even a small fraction of Edgeworth-Kuiper belt objects—all have the potential to become K-Pg-type impactors. Not only could this process have played a role in shaping the directions that life evolved on Earth, it will almost certainly impact terrestrial life in the future.

## 9. Conclusion

Centaurs, and other icy objects, can migrate from beyond Saturn and Uranus, even from the Scattered Disk, to encounter Jupiter. In our simulations Jupiter repeatedly captures Centaurs that pass into its gravitational sphere of influence into long-term encounters—even temporary captures—over a fairly wide range of initial encounter relative velocities. The magnitude and orientation of the velocity vector when the objects leave Jupiter's sphere of influence determines whether the object has been placed into a JFC orbit post-encounter. Simple statistics of successful Centaur to JFC conversions support the model presented in two variants in Figures 5 and 8: that Centaurs are converted into JFCs through a process of perihelion/aphelion interchange during extended encounters with Jupiter. Not only do our simulations suggest that Saturn can also place Centaurs into orbits with aphelia at Saturn, and perihelia interior to Jupiter, several such objects—which we dub "Saturn Family Comets" or SFCs, had already been discovered in such orbits but not recognized as a distinct cometary subfamily family until now. Given the ability of Jupiter and Saturn to place planetesimals into Earth-crossing JFC and SFC orbits, and given the result from GCH18 that the Centaurs and Scattered Disk appear to be dynamically indistinct—the Centaurs and Scattered Disk objects are all potentially K-Pg-type impactors. The impact threat to Earth from Centaurs and Scattered Disk objects is both ongoing and permanent.

**Acknowledgements:** The authors thank Philip Sharp for allowing some of the simulations presented in this study to run on the University of Auckland Math Department computer network. A portion of this work was conducted at the U.S. Military Academy, West Point, NY, and part at the Jet Propulsion Laboratory, California Institute of Technology, under contract to NASA. U.S. Government sponsorship acknowledged. The views expressed in this article are those of the authors and do not reflect the official policy or position of the Department of the Army or Department of Defense.

## References

Alvarez L. W. et al., 1980, Science, 208, 4448, 1095.

Artemieva N., Morgan J., 2009, Icarus 201, 768.

Bailey M. E., Malhotra R., Icarus, 203, 155.

Brouwer D., 1937, Astron. J., 46, 149-153.

Chodas P. W., Yeomans D. K., 1996, IAU Colloquium, 156, 1.

Collins G. S. et al., 2008, EPSL, 270, 221.

Danby J. M. A., 1988, Fundamentals of Celestial Mechanics, 2nd ed. Willmann-Bell, Richmond, VA.

Duncan M.J., Levison H.F., 1997, Science, 276, 1670.

Gandomi A., Haider M., 2015, Int. J. Information Management, 35, 137.

Grazier K.R., et al., 1999, Icarus, 140, 341.

Grazier K.R., et al., 2005a, ANZIAM J., 46(E), C1086.

Grazier K.R., et al., 2005b, ANZIAM J., 46(E), C101.

Grazier K.R., et al., 2013, Astron. J. 145, 112.

Grazier K.R., et al, 2014, Icarus, 232, 13.

Grazier K. R., Astrobiology, 16, 23.

Grazier K.R., et al., 2018, Astron. J., 156, 232.

Hammel H. B. et al., 1995, Science, 267, 1288.

Holman M. J., Wisdom J., 1993, Astron. J. 105, 1987.

Horner J., et al., 2004, MNRAS, 354, 798.

Horner J, Jones B.W., 2008, Int. J. Astrobiology, 7, 251.

Horner J, Jones B.W., 2009, Int. J. Astrobiology, 8, 75.

Horner J, et al., 2010, Int. J. Astrobiology, 9, 1.

Jablonski, D., Chaloner, W. G., 1994..Philosophical Transactions of the Royal Society of London. 344, 1307: 11.

Labrinidis A., Jagadish H. V., 2012, Proceedings of the VLDB Endowment, 5, 2032.

Lewis, A.R., 2013, *Astron. J*. 146, 16.

Levison H. F., Duncan M. J., 1997, Icarus 127, 13.

Moore J.R., Sharma, M., 2013, Proceedings of the 44th Lunar and Planetary Conference; Woodlands, TX, 2431.

Pope K.O., et al., 1997, J. Geophys. Res., 102, 21,645.

Rickman H., Malmort A. M., 1981, A&A, 102, 165.

Störmer C., 1907, Arch. Sci. Phys. Nat, 24, 5-18, 113-158, 221-247.

Tancredi G. et al., 1990, A&A, 239, 375.

Vickery, A.M., H.J. Melosh,, 1990, Global Catastrophes in Earth History; An Interdisciplinary Conference on Impacts, Volcanism, and Mass Mortality, eds V.L. Sharpton and P.D. Ward, Geological Society of America.

Volk K., Malhotra R., 2008, Astroph. J. 687, 714.

Ward, P. D., Brownlee, D., 2000, In: Rare Earth: Why Complex Life is Uncommon in the Universe, 235.

Wetherill, G.W., 1994, Astrophysics and Space Science, 212, 23.

Wetherill, G.W., 1995, Nature 373, 470.

Wilson, H.C., 1909, Popular Astronomy, 17, 629.

**Table 1**. Percentage (in %) correlations between changes in semi-major axes and eccentricity for encounters with each planet, sorted by simulation. Tabulated vertically are increases/decreases in semi-major axis; tabulated horizontally are changes in eccentricity. For example, for encounters with Jupiter in the JS simulations, 28 percent of the particles that had an increase in semi-major axis also had an increased eccentricity. For encounters with Neptune in the UN simulations, 18 percent of the particles had a decrease in semi-major axis with an increase in eccentricity.

|  |  | JS | | SU | | UN | |
|---|---|---|---|---|---|---|---|
|  | Δa/Δe | inc | dec | inc | dec | inc | dec |
| **Jupiter** | **inc** | 28 | 22 | 28 | 22 | 28 | 22 |
|  | **dec** | 23 | 27 | 23 | 27 | 23 | 27 |
| **Saturn** | **inc** | 40 | 10 | 40 | 10 | 42 | 08 |
|  | **dec** | 13 | 37 | 13 | 37 | 11 | 38 |
| **Uranus** | **inc** | 32 | 17 | 32 | 17 | 37 | 11 |
|  | **dec** | 18 | 33 | 18 | 33 | 14 | 38 |
| **Neptune** | **inc** | 31 | 18 | 31 | 18 | 32 | 16 |
|  | **dec** | 19 | 32 | 19 | 32 | 18 | 33 |

**Table 2**. Average durations in days for the encounter scenarios tabulated in Table 4. For example (using the same cells as in the previous example), for encounters with Jupiter in the JS simulations, the particles that had an increase in semi-major axis and also had an increased eccentricity had an average encounter duration of 134 days. For encounters with Neptune in the UN simulations, the particles that had a decrease in semi-major axis with an increase in eccentricity had an average encounter duration of 44899 days.

| | | JS | | SU | | UN | |
|---|---|---|---|---|---|---|---|
| | | **inc** | **dec** | **inc** | **dec** | **inc** | **dec** |
| **Jupiter** | **inc** | 134 | 523 | 2657 | 4413 | 2597 | 4195 |
| | **dec** | 477 | 135 | 4489 | 2668 | 4379 | 2564 |
| **Saturn** | **inc** | 146 | 475 | 3015 | 14946 | 2416 | 16592 |
| | **dec** | 510 | 152 | 16281 | 2920 | 18022 | 2275 |
| **Uranus** | **inc** | 192 | 326 | 6713 | 14408 | 5755 | 22231 |
| | **dec** | 323 | 200 | 15124 | 7131 | 25296 | 5406 |
| **Neptune** | **inc** | 470 | 635 | 12892 | 27120 | 17952 | 40558 |
| | **dec** | 601 | 537 | 27626 | 12864 | 44899 | 17782 |

**Figure 1**. Aphelion distance vs. perihelion distance for particles passing through the inner Solar System (q < 1.5 AU) in full-mass simulations. Red points are for orbits where the perihelia fell interior to Mars, but exterior to Earth. Blue points are for orbits interior to Earth, orange points are Venus-crossers, and black points are for objects that passed interior to Mercury. Figure is a replot/rescale of a subset of the information presented in Fig. 5b from G16.

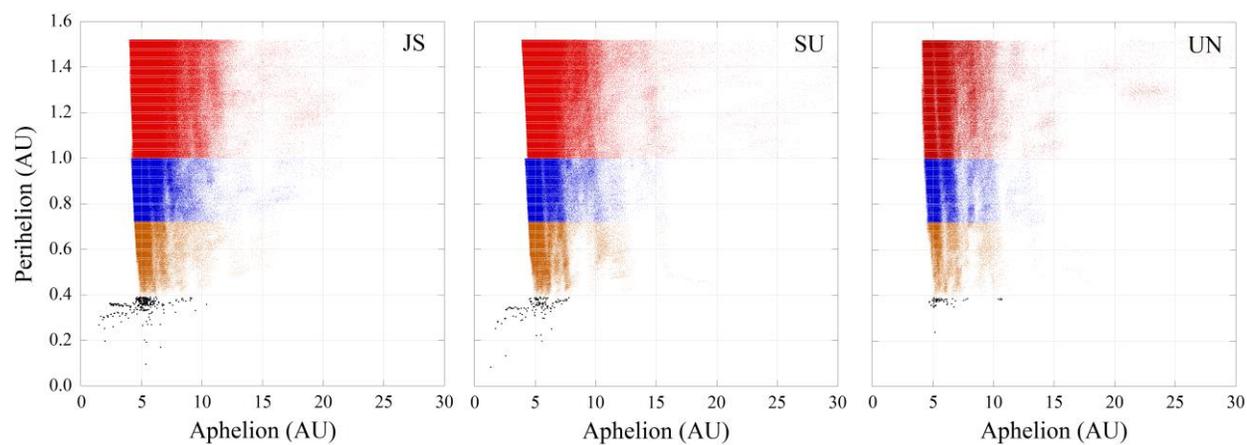

**Figure 2.** Encounter duration in days versus planetesimal/planet relative velocity at the onset of a close approach—the instant the particle enters a gravitational sphere of influence. Panel A represents encounters with Jupiter, B is Saturn, C is Uranus, D is Neptune. The plot displays encounters with particles originating in the JS (orange), SU (green/yellow), and UN (green/blue) zones. Negative duration values represent retrograde close approaches. Cyan points in Panel A represent events where a Centaur was converted to a JFC with an aphelion $Q <= 5.2$ AU and perihelion $q <= 3.3$ AU.

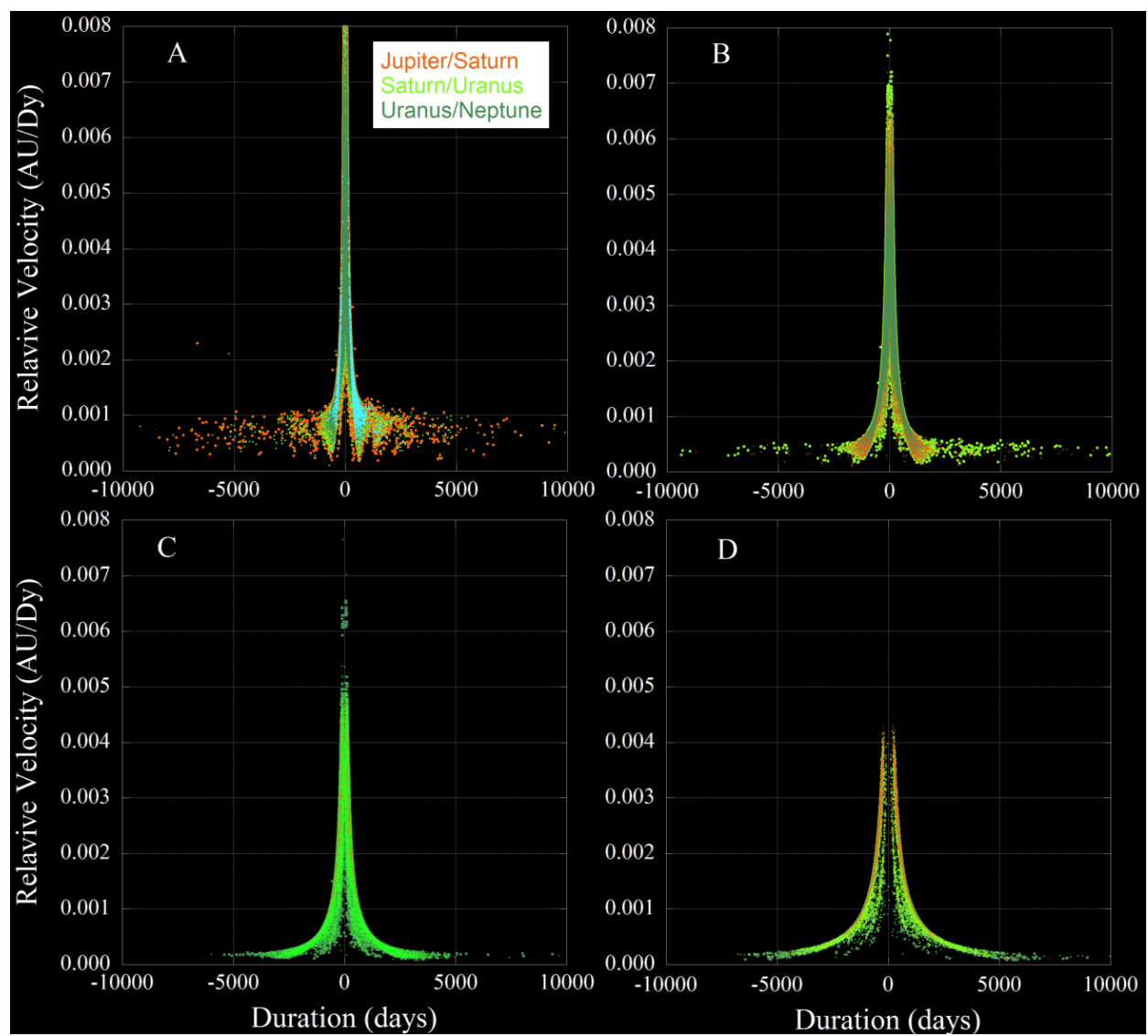

**Figure 3.** The geometry of a gravity assist trajectory—one that increases both a planetesimal's semi-major axis and eccentricity. The reference frame is sun-centered ecliptic: a rotating frame with the planet-Sun line is presumed to lie on the $-y$ axis. The vector diagram beneath shows that although the magnitudes of the inbound ($v_{in}$) and outbound ($v_{out}$) planet/planetesimal relative velocity vectors are equal, when translated into the heliocentric frame by adding the velocity of the planet (yielding $v_{HI}$ and $v_{HO}$) the velocity vector changes direction and increases in magnitude. Image Credit: NASA/JPL-Caltech/SwRI/MSSS.

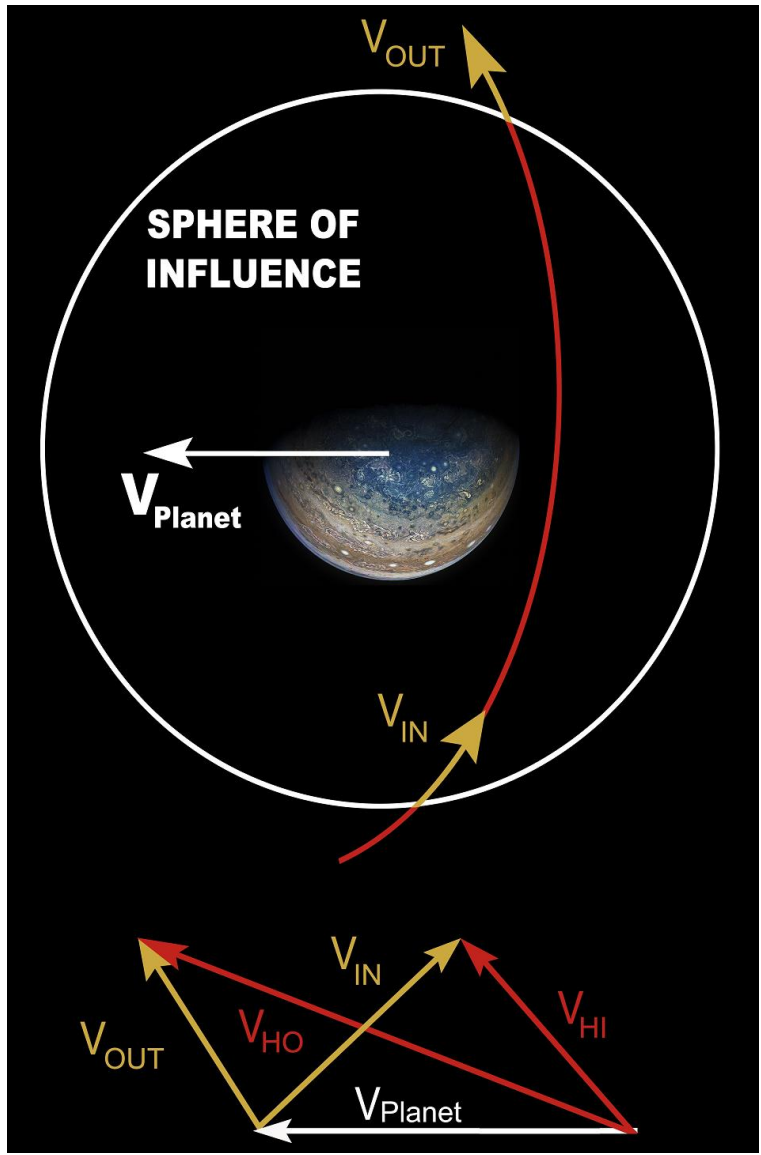

**Figure 4.** Mined geometries of the Centaur-to-JFC conversation process. Plotted in the three panels below—one for each planetesimal reservoir—are geometries for close approach events where particles enter into Jupiter's sphere of influence having perihelia greater than or equal to 5.2 au, and exit having aphelia less than or equal to 5.2 au, and perihelia less than 3.5 au. The radial units are in multiples of the radius of Jupiter's dynamical sphere of influence, and the reference frame is sun-centered ecliptic, with 0° representing the Jupiter-Sun line. Black points indicate close approach ingress positions. Red radial depict egress geometry with each line representing the number of particles in 10° bins, plotted along the centerline of each bin, and are normalized to 1.0.

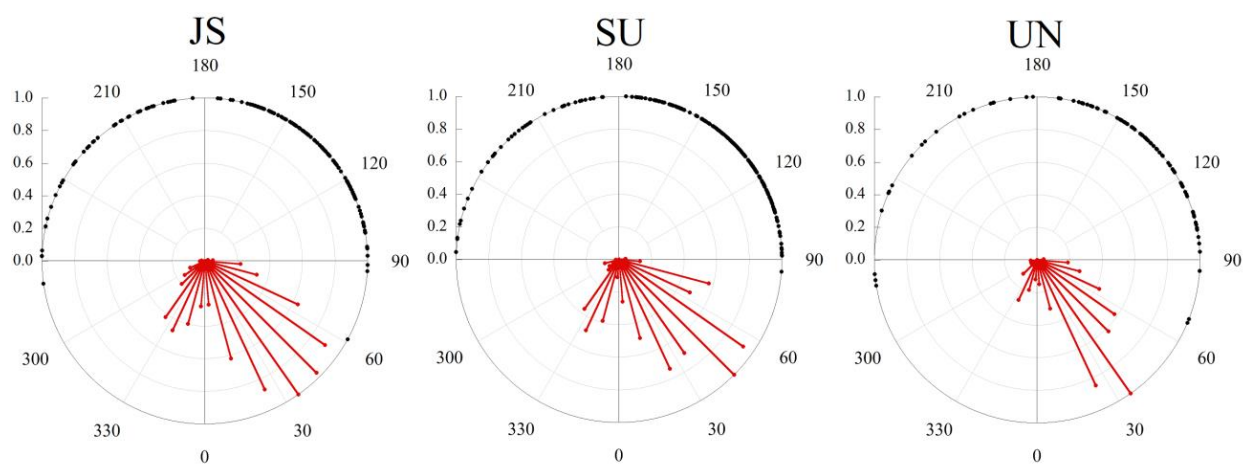

**Figure 5.** An idealized illustration of the process by which a Centaur is converted into a Jupiter Family Comet, based upon the output displayed in Fig. 4. The planetesimal encounters Jupiter's sphere of influence at, or near, perihelion., undergoes a high-deflection encounter—or even a long-term temporary capture, performing at least one full orbit of Jupiter—and exits as shown. In the vector diagram describing the encounter beneath, particles near their perihelion would overtake Jupiter, and would be moving nearly parallel to Jupiter upon exit from the close approach, but much slower in the heliocentric frame. This fixes the point where the particle leaves Jupiter's sphere of influence as the particle's new aphelion. The vector diagram beneath shows how $v_{HI} < v_{HO}$, leaving the particle in a more tightly-bound orbit (or $\Delta a < 0$). Image Credit: NASA/JPL-Caltech/SwRI/MSSS.

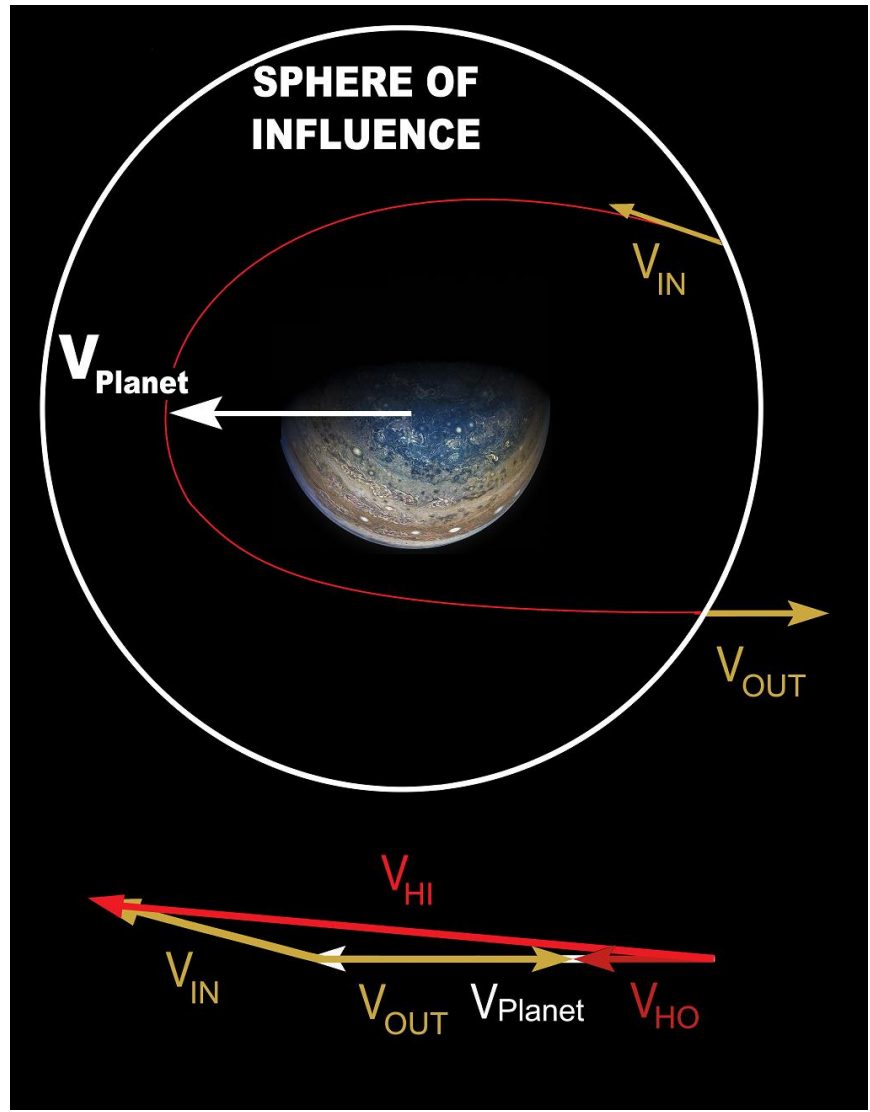

**Figure 6.** Ingress/egress geometries for all three simulation zones for close approach events where particles enter into Saturn's sphere of influence having perihelia greater than or equal to 5.2 au, and exit having aphelia less than or equal to 9.6 au, and perihelia less than 5.2 au. The radial units are in multiples of the radius of Saturn's dynamical sphere of influence, and the reference frame is sun-centered ecliptic, with 0° representing the Saturn-Sun line. Black points indicate close approach ingress positions. Red radial depict egress geometry with each line representing the number of particles in 10° bins, plotted along the centerline of each bin, and are normalized to 1.0.

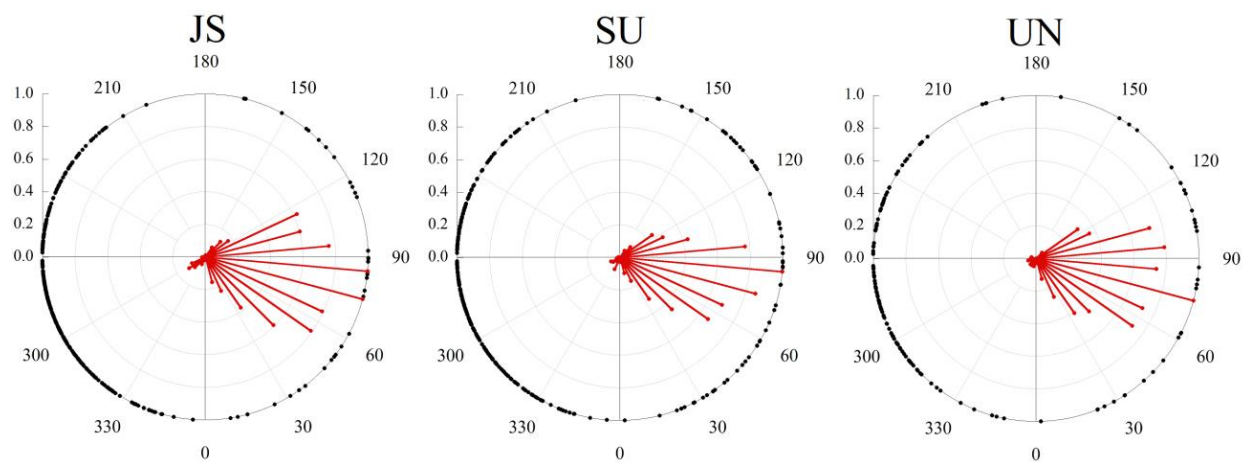

**Table 3.** Jovian Family Comets: candidate SFCs, UFCs, and NFCs. Entries from the JPL *Horizons* database of objects with aphelia near the orbit of Saturn (within the radius of Saturn's gravitational sphere of influence). The objects P/2005 S2 (Skiff) and 2016 EX would be classified as Centaurs, but the remainder of the objects, with perihelia in the inner Solar System, could be considered "Saturn Family Comets".

| Saturn Family Comets | | | | | |
|---|---|---|---|---|---|
| **Full Name** | **a** | **e** | **q** | **Q** | **i** |
| 944 Hidalgo (1920 HZ) | 5.74 | 0.661 | 1.95 | 9.54 | 42.52 |
| 271P/van Houten-Lemmon | 6.97 | 0.390 | 4.25 | 9.69 | 6.86 |
| 126P/IRAS | 5.65 | 0.696 | 1.72 | 9.58 | 45.80 |
| P/2001 H5 (NEAT) | 5.99 | 0.600 | 2.40 | 9.59 | 8.40 |
| P/2005 S2 (Skiff) | 7.96 | 0.197 | 6.40 | 9.53 | 3.14 |
| P/2006 R1 (Siding Spring) | 5.61 | 0.702 | 1.67 | 9.56 | 160.01 |
| P/2006 S4 (Christensen) | 6.24 | 0.508 | 3.07 | 9.41 | 39.63 |
| P/2008 L2 (Hill) | 6.00 | 0.614 | 2.32 | 9.68 | 25.86 |
| (2009 DP2) | 6.68 | 0.422 | 3.86 | 9.50 | 27.01 |
| P/2010 WK (LINEAR) | 5.73 | 0.692 | 1.77 | 9.70 | 11.48 |
| (2011 RC17) | 6.29 | 0.536 | 2.92 | 9.65 | 11.33 |
| (2011 SQ249) | 6.61 | 0.453 | 3.62 | 9.60 | 16.68 |
| P/2013 T1 (PANSTARRS) | 5.87 | 0.623 | 2.21 | 9.52 | 24.21 |
| (2015 BW524) | 7.14 | 0.330 | 4.78 | 9.49 | 9.23 |
| P/2015 PD229 (Cameron-ISON) | 7.18 | 0.327 | 4.83 | 9.52 | 2.03 |
| P/2015 P4 (PANSTARRS) | 6.07 | 0.584 | 2.53 | 9.62 | 8.71 |
| C/2015 R1 (PANSTARRS) | 5.90 | 0.633 | 2.17 | 9.63 | 22.67 |
| (2016 AF67) | 6.91 | 0.400 | 4.15 | 9.67 | 15.27 |
| 494158 (2016 EX) | 7.77 | 0.231 | 5.98 | 9.57 | 6.28 |
| "Uranus Family" Centaurs/Comets | | | | | |
| **Full Name** | **a** | **e** | **q** | **Q** | **i** |
| 2008 FC76 | 14.66 | 0.307 | 10.17 | 19.16 | 27.15 |
| 2000 DG8 | 10.76 | 0.794 | 2.21 | 19.31 | 129.32 |
| 2004 CJ39 | 12.88 | 0.477 | 6.73 | 19.02 | 3.61 |
| 2008 UZ331 | 18.36 | 0.035 | 17.71 | 19.01 | 32.63 |
| 2012 GM12 | 17.19 | 0.102 | 15.43 | 18.94 | 12.57 |
| 2014 KR101 | 14.79 | 0.285 | 10.57 | 19.02 | 9.12 |
| 2015 BG518 | 14.67 | 0.307 | 10.17 | 19.17 | 1.82 |
| 166P | 13.88 | 0.383 | 8.56 | 19.20 | 15.37 |
| "Neptune Family" Centaurs/Comets | | | | | |
| **Full Name** | **a** | **e** | **q** | **Q** | **i** |
| 330836 Orius (2009 HW77) | 21.57 | 0.421 | 12.49 | 30.65 | 17.86 |
| 427507 (2002 DH5) | 22.14 | 0.365 | 14.05 | 30.23 | 22.46 |
| 463663 (2014 HY123) | 18.82 | 0.629 | 6.98 | 30.67 | 13.93 |
| 2003 QD112 | 18.97 | 0.583 | 7.91 | 30.04 | 14.51 |
| 2007 BP102 | 23.99 | 0.260 | 17.74 | 30.23 | 64.68 |
| 2007 TJ422 | 19.46 | 0.527 | 9.20 | 29.72 | 2.91 |
| 2010 LO33 | 23.02 | 0.317 | 15.72 | 30.32 | 17.84 |
| 2012 PD26 | 20.35 | 0.505 | 10.08 | 30.62 | 7.70 |
| 2013 EZ27 | 19.73 | 0.550 | 8.87 | 30.58 | 14.61 |
| 2013 LG29 | 16.93 | 0.790 | 3.55 | 30.31 | 15.40 |
| 2015 BD518 | 23.38 | 0.304 | 16.28 | 30.49 | 17.17 |
| 2016 GC241 | 21.74 | 0.365 | 13.81 | 29.66 | 4.19 |
| 20D/Westphal | 15.64 | 0.920 | 1.25 | 30.03 | 40.89 |
| C/2002 A1 (LINEAR) | 17.16 | 0.725 | 4.71 | 29.60 | 14.05 |
| C/2002 A2 (LINEAR) | 17.19 | 0.726 | 4.71 | 29.67 | 14.05 |
| C/2012 H2 (McNaught) | 16.14 | 0.894 | 1.72 | 30.57 | 92.84 |
| C/2015 F5 (SWAN-Xingming) | 15.47 | 0.978 | 0.35 | 30.60 | 149.26 |
| C/2015 GX (PANSTARRS) | 16.19 | 0.878 | 1.97 | 30.40 | 90.25 |
| C/2015 X2 (Catalina) | 15.75 | 0.879 | 1.90 | 29.60 | 72.46 |
| C/2017 U5 (PANSTARRS) | 16.94 | 0.745 | 4.33 | 29.55 | 18.96 |

**Figure 7.** Similar to Fig. 1, figure is a replot/rescale of a subset of the information presented in Fig. 5a from G16. Displayed are the Aphelion distance vs. perihelion distance for particles passing through the inner Solar System (q < 1.5 AU) in embryo simulations where Jupiter and Saturn are 15 Earth masses. Red points are for orbits where the perihelila fell interior to Mars, but exterior to Earth. Blue points are for orbits interior to Earth, orange points are Venus-crossers, and black points are for objects that passed interior to Mercury. The tendril-like structures are due to successions of encounters with Jupiter where particles were driven ever-deeper into the inner Solar System.

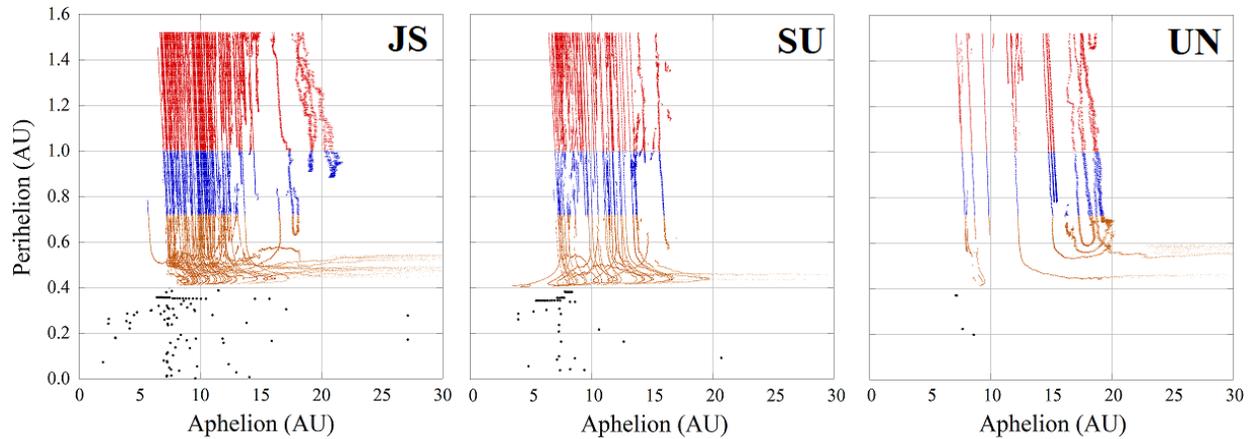

**Figure 8**. An idealized illustration of another manifestation of the process by which a Centaur is converted into a Jupiter or Saturn Family Comet, or how a planetesimal, with a perihelion interior to Jupiter, can encounter Jupiter and have its perihelion driven sunward. In this image, the Sun is towards the bottom of the page. The planetesimal encounters Jupiter's sphere of influence, undergoes a high-deflection encounter—or even a long-term temporary capture, performing at least one full orbit of Jupiter—and exits as shown. In the vector diagram describing the encounter beneath, the particle enters Jupiter's sphere of influence at any point, and would be moving nearly parallel to Jupiter upon exit from the close approach, but much slower in the heliocentric frame. This fixes the point where the particle leaves Jupiter's sphere of influence as the particle's new aphelion. The vector diagram beneath shows how $v_{HI} < v_{HO}$, leaving the particle in a more tightly-bound orbit (or $\Delta a < 0$). Image Credit: NASA/JPL-Caltech/SwRI/MSSS.

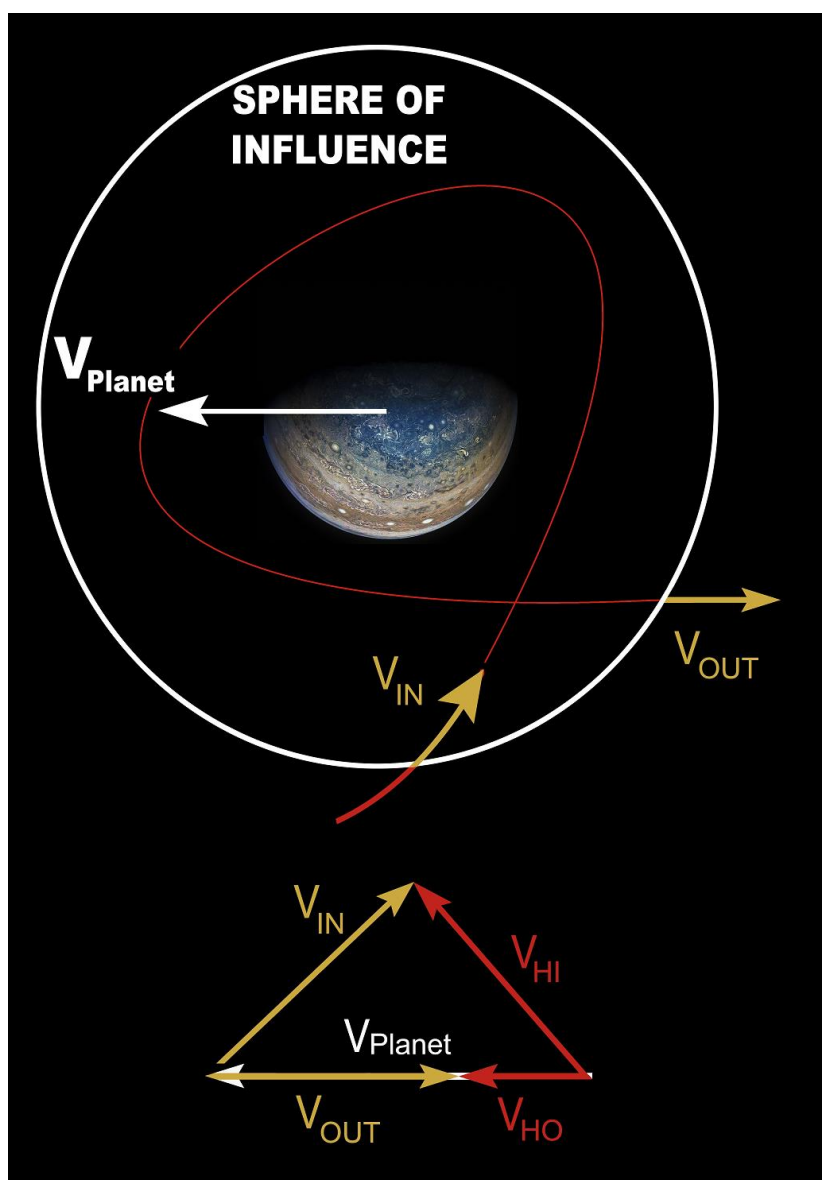